\documentclass[prl,aps,amsmath,10pt,twocolumn,showpacs]{revtex4-1} 

\usepackage{amsmath}
\usepackage{mathrsfs}
\usepackage{amsfonts}
\usepackage{graphicx}

\begin{document}

\title{Analytical prediction for the optical matrix}

\author{V. Dom\'inguez-Rocha}
\email{vdr@fis.unam.mx, vidomr@gmail.com}
\affiliation{Instituto de Ciencias F\'isicas, Universidad Nacional Aut\'onoma 
de M\'exico, Apartado Postal 48-3, 62210 Cuernavaca, Mor., Mexico}

\author{R. A. M\'endez-S\'anchez}
\email{mendez@icf.unam.mx}
\affiliation{Instituto de Ciencias F\'isicas, Universidad Nacional Aut\'onoma de M\'exico, Apartado Postal 48-3, 62210 Cuernavaca, Mor., Mexico}

\author{M. Mart\'inez-Mares}
\email{moi@xanum.uam.mx}
\affiliation{Departamento de F\'isica, Universidad Aut\'onoma 
Metropolitana-Iztapalapa, Apartado Postal 55-534, 09340 Ciudad de M\'exico, 
Mexico}

\author{A. Robledo}
\email{robledo@fisica.unam.mx}
\affiliation{Instituto de F\'isica and Centro de Ciencias de la Complejidad, Universidad Nacional Aut\'onoma
de M\'exico, Apartado Postal 20-364, 01000 Ciudad de M\'exico, 
Mexico}

\begin{abstract}
Contrary to praxis, we provide an analytical expression, for a physical locally 
periodic structure, of the average $\langle S\rangle$ of the scattering matrix, 
called optical $S$ matrix in the nuclear physics jargon, and fundamentally 
present in all scattering processes. This is done with the help of a strictly 
analogous nonlinear dynamical mapping where iteration time is the number $N$ of 
scatterers. The ergodic property of chaotic attractors implies the existence and 
analyticity of $\langle S\rangle$. We find that the optical $S$ matrix depends 
only on the transport properties of a single cell, and that the Poisson kernel 
is the distribution of the scattering matrix $S_N$ in the large size limit 
$N\rightarrow \infty$. The theoretical distribution shows perfect agreement with 
numerical results for a chain of delta potentials. A consequence of our findings 
is the a priori knowledge of $\langle S\rangle$ without resort to experimental 
data.
\end{abstract}


\maketitle

Scattering is an outstanding phenomenon of nature concerning waves and 
particles~\cite{NewtonBook}. It is important in almost all branches of physics 
since many physical observables, at microscopic and macroscopic scales, are 
described in terms of scattering 
properties~\cite{Krane,MelloBook,Landauer,Buttiker1,Buttiker2,Brouwer1,Brouwer2,
Doron,Schanze1,Mendez,Hemmady,FloresO}. In experiments, the scattering 
amplitudes vary with respect to a tuning parameter which could be the energy of 
incidence in nuclear, many body, and atomic 
physics~\cite{Orrigo,Watson,Silvestri}, the Fermi energy or an applied magnetic 
field in condensed matter physics~\cite{Keller,Marcus}, and the frequency in 
optics~\cite{Tribelsky}, microwaves~\cite{Schanze1,Mendez,Kuhl,Schanze2,Sirko}, 
and elastic systems~\cite{Marcel,FloresO}.

Irrespective of the different mechanisms that occur in complex scattering 
processes, the wave amplitudes contain a direct (rapid) component and an 
equilibrated (delayed) multiple-scattering component. The rapid response is 
revealed by the average of the measured scattering amplitudes over an interval 
of the corresponding tuning parameter, the so called optical 
amplitudes~\cite{LMS}. The delayed response is obtained by subtracting the 
direct response to the set of measured scattering amplitudes. The scattering 
matrix samples amplitude values according to a range or space fixed by the 
current value of the tuning parameter. The way in which the values of the 
scattering matrix are distributed along this space is determined once the 
optical $S$ matrix is specified, any other information being 
irrelevant~\cite{MelloLesHouches}. 

The scattering amplitudes fluctuate around their average values or from 
sample-to-sample, so that, in a statistical-mechanical fashion, the construction 
of an ensemble of scattering systems together with an ergodic hypothesis is used 
to obtain a quantitative description of the scattering process. The direct 
response component is corroborated via comparison with the ensemble average of 
the scattering amplitudes at a fixed tuning parameter~\cite{LMS,MPS}. In all 
cases $\langle S\rangle$ is determined from the experimental or numerical data. 
The question we address here is whether there exists the possibility to predict 
the matrix $\langle S\rangle$ via a procedure that does not require obtaining 
the average from the experiment, actual or numeric, but follows instead from a 
scattering formalism.

\begin{figure}
\includegraphics[width=\columnwidth]{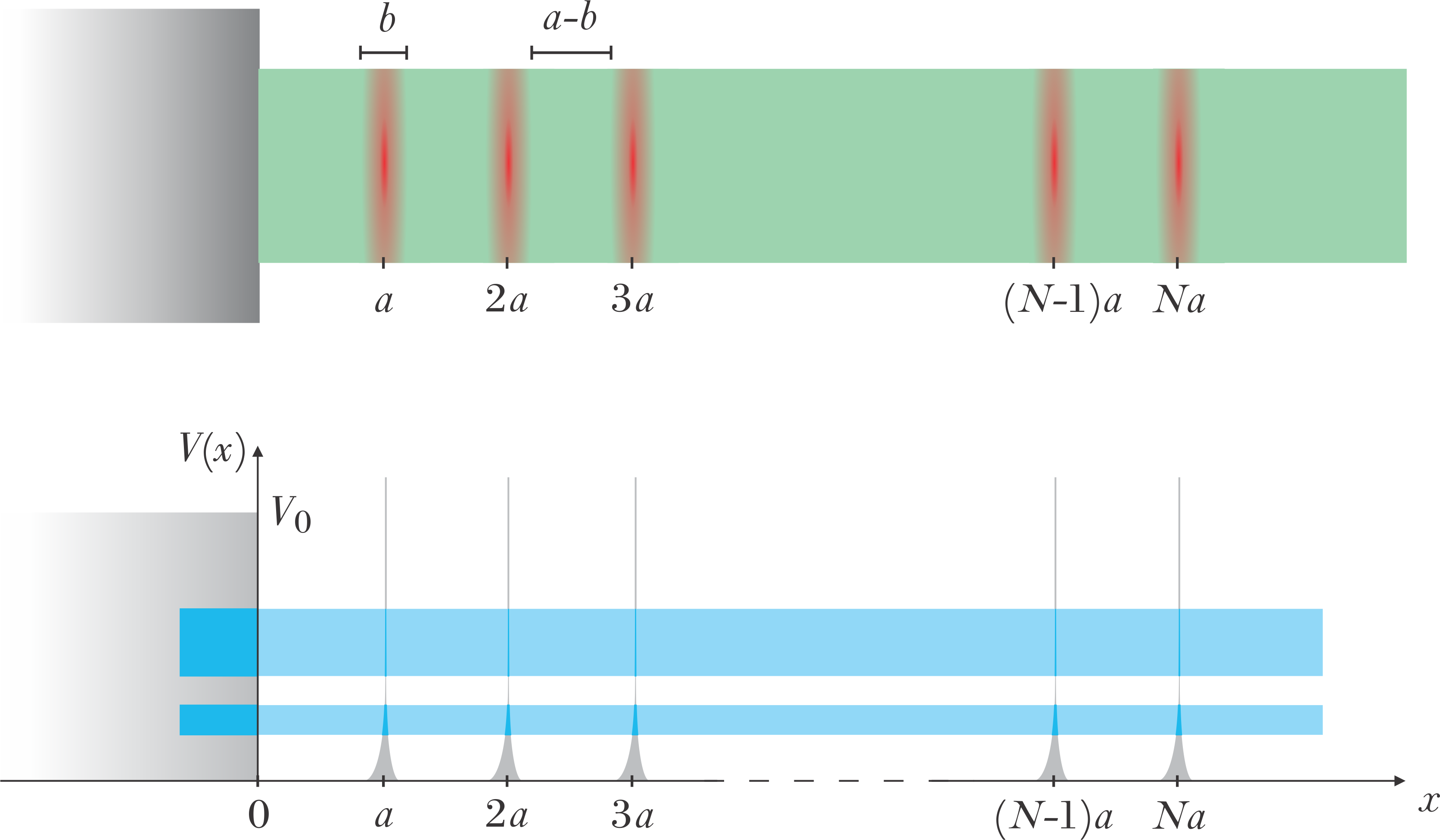}
\caption{(Color online) The upper figure illustrates a locally periodic system 
composed by two media, blocked at the left. It represents an elastic system,
a microwave waveguide, a one-dimensional photonic crystal, or a superlattice.
The lower figure is a schematic representation of the potential function in the 
corresponding quantum case, where the first sub-bands are shown as blue 
fringes.}
\label{fig:Fig1}
\end{figure}

Here, we provide a physical multiple-scattering setting that leads to an 
exact analytical expression for the optical $S$ matrix. This expression appears 
in the thermodynamic or large size limit as the number of scatterers 
$N\rightarrow \infty$. For locally periodic structures the scattering process 
has been shown to be analogous to a nonlinear dynamical evolution such that the 
scattering matrix gradual change with size is equivalent to iteration of a 
dissipative mapping that displays transitions to chaos of the tangent 
bifurcation type~\cite{MR,VictorJOPA,MDR}. The transitions from localized to 
extended states correspond to transitions from regularity to 
chaos~\cite{MR,VictorJOPA,MDR} and the chaotic regimes hold an ergodic 
property~\cite{Schuster}.

\begin{figure}
\includegraphics[width=\columnwidth]{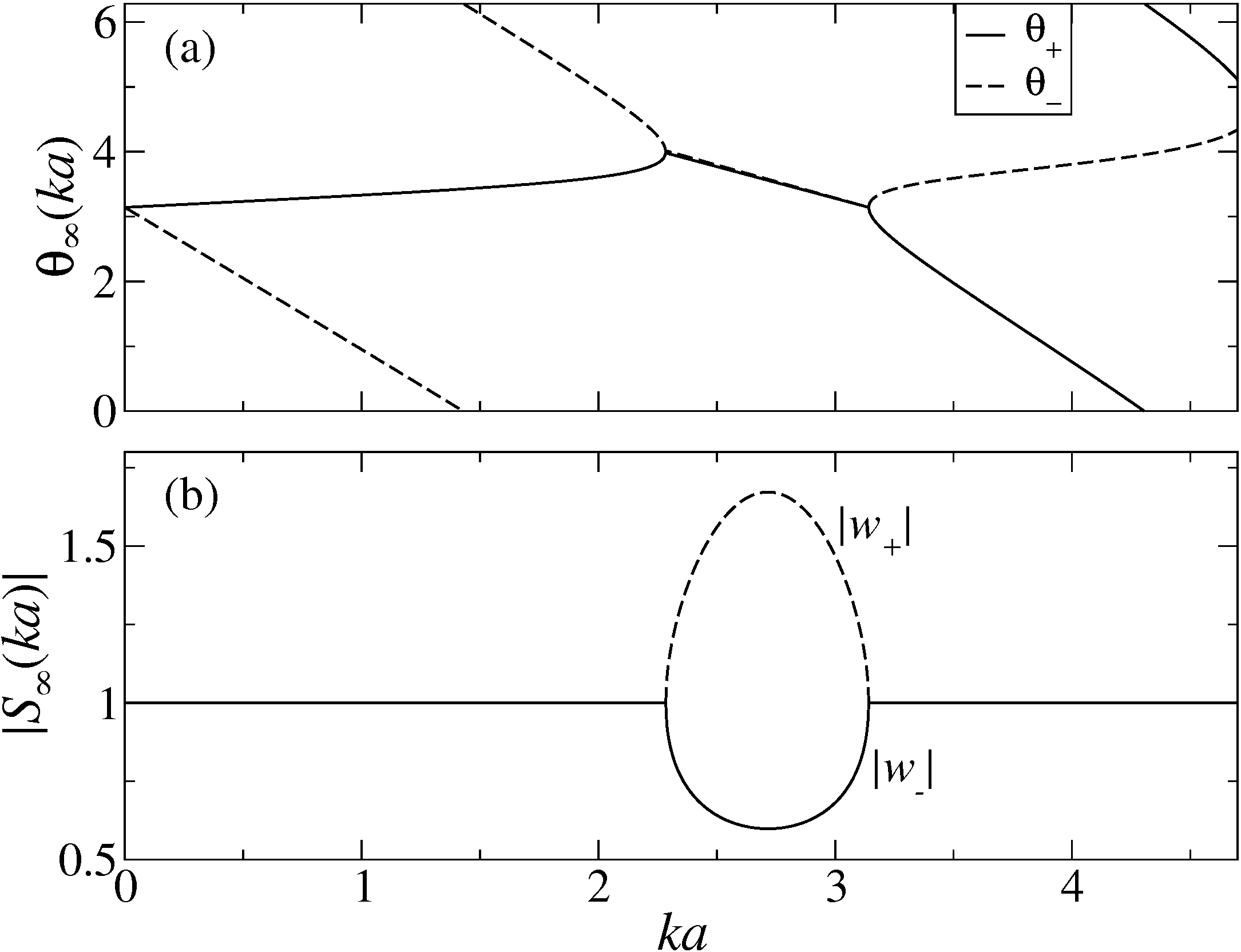}
\caption{\footnotesize 
Fixed point solutions $S_{\infty}(ka)$,
Eqs.~(\ref{eq:Sinf})-(\ref{eq:w}), (a) phase and (b) modulus. The phases
of $w_{+}$ and $w_{-}$ are the same within the gaps but the stable solutions are
$\theta_{+}(ka)$ or $\theta_{-}(ka)$ depending on the gap defined by the values 
of $ka$. Note that $|w_+(ka)|>1$ while $|w_{-}(ka)|<1$ within the band.}
\label{fig:Fig2}
\end{figure}

Let us consider a single port quantum system in one dimension, as shown in 
Fig.~\ref{fig:Fig1}; it consists of a periodic array with a finite number of 
identical scattering elements~\cite{Griffiths,Morales,Luna}. The system with 
$N-1$ scatterers is described by the $1\times 1$ scattering matrix $S_{N-1}$, 
and this can be immediately related to the scattering matrix $S_N$ of a system 
with $N$ scatterers by addition of another scatterer. The result is the 
recurrence relation ~\cite{VictorJOPA}
\begin{equation}
\label{eq:recursive}
S_N = \frac{r'_bz_b^* + z_bS_{N-1}}
{{r'_b}^*z_b + z_b^*S^*_{N-1}}S^*_{N-1},
\end{equation}
where $r_b$ ($t_b$) and $r'_b$ ($t'_b$) are the reflection (transmission) 
amplitudes of an individual scatterer for incidence on the left and right, 
respectively; 
$z_b=t_b\mathrm{e}^{\mathrm{i}\phi/2}\mathrm{e}^{\mathrm{i}k(a-b)}$, with 
$\mathrm{e}^{\mathrm{i} \phi/2}=t'_b/t_b$, $k$ is the incident wave number, $b$ 
the width of the scatterer, and $a$ the lattice constant. Since $S_N$ depends 
intrinsically on $k$, a gap and band structure emerges with respect to $k$ as 
$N$ increases (see below). The gaps and bands are clearly formed in the 
thermodynamic limit $N\rightarrow \infty$, for which Eq.~(\ref{eq:recursive}) 
has a stable and an unstable fixed point solutions for each value of $k$; these 
are~\cite{VictorJOPA}
\begin{equation}
\label{eq:Sinf}
S_{\infty}(k) = \left\{ 
\begin{array}{ll}
\mathrm{e}^{\mathrm{i}\theta_\pm(k)}, & 
k\in\, \mbox{gap} \\ 
w_\pm(k), & k\in\, \mbox{band},
\end{array}
\right. \, 
\end{equation}
where 
\begin{equation}
\mathrm{e}^{\mathrm{i}\theta_\pm(k)} =  
\frac{\pm\sqrt{\left[\mathrm{Re}\, z_b(k)\right]^2- 
\left| t_b(k) \right|^4}+ 
\mathrm{i}\, \mathrm{Im}\, z_b(k)}
{r_b^*(k)z_b(k)},
\label{eq:e}
\end{equation}
and 
\begin{equation}
\label{eq:w}
w_{\pm}(k) = \mathrm{i}\, 
\frac{\pm\sqrt{ \left| t_b(k) \right|^4-\left[\mathrm{Re}\, 
z_b(k)\right]^2} + 
\mathrm{Im}\, z_b(k)}{r_b^*(k) z_b(k)} \, .
\end{equation}
These solutions are shown in Fig.~\ref{fig:Fig2}.

\begin{figure}
\includegraphics[width=\columnwidth]{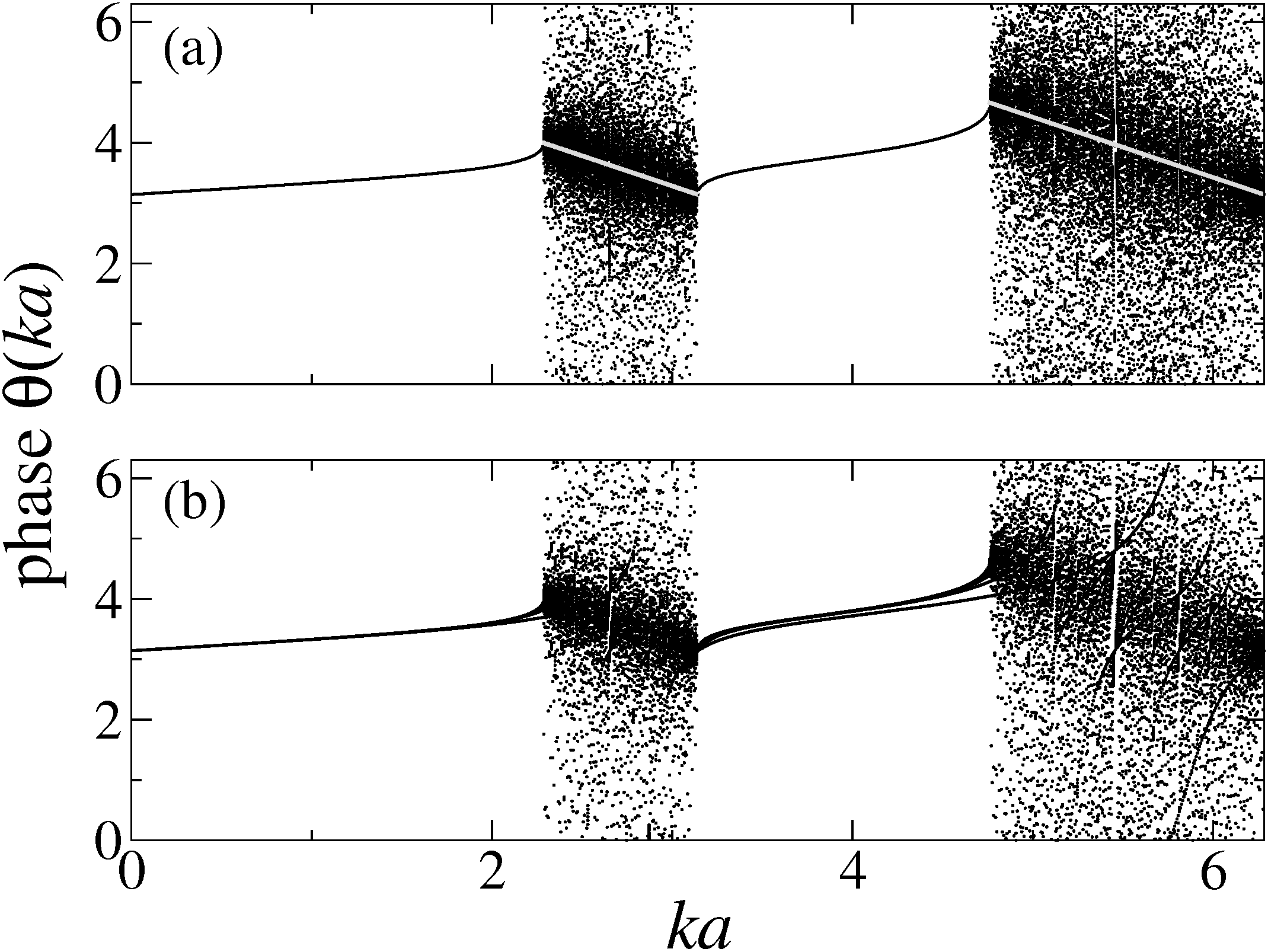}
\caption{\footnotesize Phase $\theta$ of the scattering matrix as a function of 
incident energy $ka$. In (a) we plot the last 30 of 1000 iterations of the 
recurrence relation Eq.~(\ref{eq:recursive}) and also the analytical expression 
for the stable solution of Eq.~(\ref{eq:Sinf}) (see Fig.~\ref{fig:Fig2}). The 
stable solution is indistinguishable from the numerical result within the gaps, 
while it gives the maximum of the distribution of the phase within the bands 
(light line). See Fig.~\ref{fig:Fig4}. In (b) we show the evolution of the 
phase from $N=1$ up to $N=30$, where the additional black lines are transient 
values associated with the unstable (repellor) solutions.}
\label{fig:Fig3}
\end{figure}

The recurrence relation Eq.~(\ref{eq:recursive}) for the (complex number) 
scattering matrix $S_{N}$ is also a recurrence relation for the phase 
$\theta_{N}$, $S_N=\exp(\mathrm{i}\theta_N)$, which can be interpreted as a 
real number nonlinear iterated map. The main features of this map, including 
its generalization to Bethe lattice arrangements of scatterers, have been 
determined in Refs.~\cite{MR,VictorJOPA,MDR}. It was found that the bifurcation 
diagram, or families of attractors ordered according to the control parameter 
$ka$, is made of intervals of regular attractors of period 1 separated by 
sectors of chaotic attractors. The attractors at the boundaries between the two 
kinds of families are transitions to chaos of the tangent bifurcation 
type~\cite{Schuster}. The chaotic attractors in the vicinity of the tangent 
bifurcations exhibit intermittency of type I~\cite{Schuster}. Therefore the gap 
and band behavior of the scattering system in Fig.~\ref{fig:Fig3}(a) is 
actually the map bifurcation diagram in the language of the dynamical system 
analog.

When $N\rightarrow \infty$ the value $S_{\infty}$ of the scattering matrix is 
fixed within the gaps and given by the complex number of modulus 1 in 
Eq.~(\ref{eq:Sinf}). The value of the scattering matrix fluctuates 
within the bands, but, as we shall conclude shortly, its average value $\langle 
S\rangle$ is fixed by the non unitary complex number $S_{\infty}$ in 
Eq.~(\ref{eq:Sinf}), see Fig.~\ref{fig:Fig2}. From Eq.~(\ref{eq:recursive}) we 
observe that 
\begin{equation}
\label{eq:AC1}
\lim_{N\to\infty} S_N^m= \left( \lim_{N\to\infty}S_N\right)^m = S_{\infty}^m, 
\end{equation}
with $m$ an integer. This property coincides with the analyticity condition 
satisfied by the average of the scattering matrix~\cite{MelloLesHouches}. The 
scattering matrix visits its available space according to a certain distribution 
fixed by the value of $ka$ within the bands. This distribution becomes specified 
since the analyticity condition, expressed in Eq.~(\ref{eq:AC1}), implies that 
all of its moments are known; the resulting distribution is given by Poisson's 
kernel~\cite{MelloBook,LMS,MelloLesHouches}, namely
\begin{equation}
\label{eq:Poisson}
p_{\langle S\rangle}(S) = \frac{1}{2\pi}\, 
\frac{ \left| 1- \left|\left\langle S\right\rangle\right|^2\right|}
{\left|S-\left\langle S\right\rangle\right|^2}, \quad \mbox{with}\quad 
\langle S\rangle = S_\infty,
\end{equation}
where the absolute value in the numerator was added to ensure a positive defined 
distribution for the case of an over unitary value 
$\langle S\rangle=w_{+}(ka)$. Thus, $p_{\langle S\rangle}(S)$ depends on $k$ 
through the reflection and transmission amplitudes of a single scattering 
element, Eqs.~(\ref{eq:e}), and~(\ref{eq:w}).

The results in Eqs.~(\ref{eq:AC1}) and (\ref{eq:Poisson}) are 
anticipated~\cite{MelloBook,LMS,MelloLesHouches} to be of general validity and 
can be corroborated by considering the nonlinear map analogs of the scattering 
processes described here and in Refs.~\cite{MR,VictorJOPA,MDR}. The chaotic 
dynamics within the bands formed by these model systems are ergodic and ensure 
the existence of an invariant distribution $P(\theta)$ (actually, the 
distribution for the phase $\theta$). That is, iteration time averages for map 
trajectories within the chaotic attractors are reproduced by ensemble averages. 
The identification of $p_{\langle S\rangle}(S)$ with the Poisson kernel and of 
$S_{\infty}$ with $\langle S\rangle$ is accomplished by determination of 
$P(\theta)$ and its average.

The quantum chain of delta potentials shown in the lower part of 
Fig.~\ref{fig:Fig1} will be used as a specific example \cite{VictorJOPA}. To 
obtain a $1\times 1$ scattering matrix it is necessary that energy of the 
particle is smaller than the height $V_0$ of the step potential on the left side 
of the chain. In this case $r_b=r'_b=-u/(u-2\mathrm{i}k)$ and 
$t_b=t'_b=-2\mathrm{i}k/(u-2\mathrm{i}k)$, with $u$ the intensity of the delta 
potential in the same units as the wave number. We take the initial condition 
$S_0=\mathrm{e}^{\mathrm{i}\pi}$, which corresponds to $V_0\to\infty$. In 
Fig.~\ref{fig:Fig3} the phase of the scattering matrix is plotted for $ua=10$ 
as a function of $ka$ (dimensionless quantities are used). In 
Fig.~\ref{fig:Fig3}(a) the last 30 of 1000 iterations are plotted, as well as 
the fixed point solution (we show only the stable solution) which is highlighted 
in the bands but it is indistinguishable from the numerical result within the 
gaps. In Fig.~\ref{fig:Fig3}(b) the first 30 iterations show the distribution 
of points around a maximum value given by the fixed point solution. 

\begin{figure}
\centering
\includegraphics[width=\columnwidth]{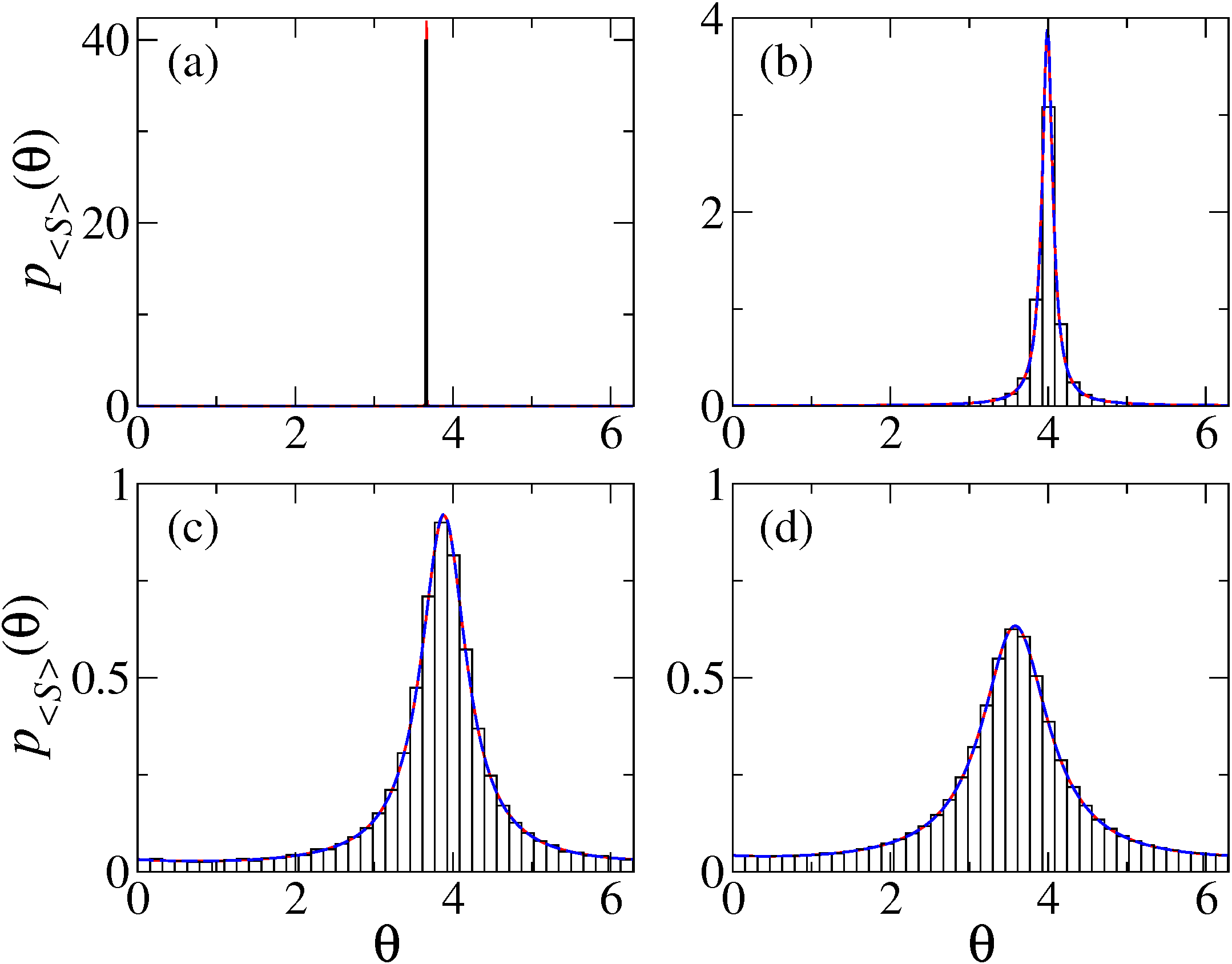}
\caption{\footnotesize Distribution of the phase for $N=10^4$ iterations of the
recurrence relation, Eq.~(\ref{eq:recursive}), for a chain of delta potentials 
with intensity $ua=10.0$ and initial condition $\theta_0=\pi$. The histograms 
are obtained at fixed $k$: (a) $ka=2.1$, (b) $ka=2.29$, (c) $ka=2.4$, and (d) 
$ka=2.7$. The dashed lines (blue lines) correspond to the theoretical model of 
Eq.~(\ref{eq:Poisson}) in which the average is given by 
$\langle S\rangle=S_\infty$. Although we show only results for the stable fixed 
point solutions $w_{-}$ the same theoretical distribution holds for the 
unstable solution $w_{+}$. The continuous lines (red lines) also correspond to 
Eq.~(\ref{eq:Poisson}) but $\langle S\rangle$ is obtained from the numerical 
data.}
\label{fig:Fig4}
\end{figure}

The numerically-obtained distributions for the phase $P(\theta)$ from an 
ensemble of $N=10^4$ scattering systems are shown as histograms in 
Fig.~\ref{fig:Fig4} for several values of $k$. In Fig.~\ref{fig:Fig4}(a) we 
observe that, for $k$ within the first gap, the distribution is just a delta 
function centered at the stable fixed point solution. The fixed point solution 
is exponentially reached as $N$ approaches the limit 
$N\to\infty$~\cite{VictorJOPA}. A similar outcome happens for the unstable fixed 
point solution in which the phase remains there for all $N$. In 
contradistinction, for $k$ inside, starting close to the border of the band and 
progressively at values further inside the band, panels (b), (c) and (d) of 
Fig.~\ref{fig:Fig4}, respectively, the phase $\theta$ of $S$ becomes 
distributed in the complete space between 0 and $2\pi$. Near the transition from 
gap to band the distribution is narrower than that for $k$ close to the center 
of the band. At the gap/band transition the fixed point solution is reached as 
a power law~\cite{VictorJOPA}. In all of the cases shown, the histograms have 
an excellent agreement with the Poisson kernel (\ref{eq:Poisson}) for $\langle 
S\rangle$ calculated from Eqs.~(\ref{eq:Sinf}) with $S_\infty=\langle S\rangle$, 
as well as when it is calculated from the numerical data. This excellent 
agreement shows that $S_{\infty}$ is the optical $S$ matrix. 

In conclusion, an analytical expression for the average of the scattering 
matrix, known as the optical $S$ matrix in the nuclear physics terminology, was 
calculated for the first time. This was done by demonstrating that the fixed 
point solution $N\rightarrow\infty$ of the size recurrence relation of the 
scattering matrix, for a locally periodic system, satisfies the analyticity 
condition. When $N\rightarrow\infty$ a perfect crystal, with consecutive gaps 
and bands, or localized and extended states, is obtained as the incident energy 
$ka$ is increased. There is a precise analog of the scattering problem with a 
nonlinear (dissipative) iterated map, such that the gap and bands of the 
scattering system correspond, respectively, to the regular and chaotic 
attractors in the nonlinear dynamics. The boundaries between corresponding to 
transition from localized to extended states can be identified with transitions 
in or out of chaos of the tangent bifurcation kind. A complete 
transcription from the two languages can be established, such that, for 
instance, the vanishing of the Lyapunov exponent at the transitions to chaos 
corresponds to the divergence of the localization length at the gap to band 
boundaries. The nonlinear map exhibits two types of dynamical properties: those 
of trajectories towards the attractor and those inside the attractor. In the 
former case the number of iterations corresponds to the number of scatterers 
$N$, and therefore describes system size growth. In the latter case map 
iterations have a different meaning. For a regular attractor, period one in our 
example, iterations only repeat certain values of the map variable, in our 
example iterations keep the phase and the scattering matrix at their constant 
fixed-point values, $\theta_{\infty}$ and 
$S_{\infty}=\exp(\mathrm{i}\theta_{\infty})$. For the chaotic attractors 
iterations change repeatedly the value of the map variable sampling the space 
spanned by the attractor, in our example the phase $\theta$ interval $[0,2\pi]$, 
according to a given distribution, the invariant distribution. The chaotic 
attractors that appear for given intervals of the control parameter $k$ possess 
the ergodic property, such that iteration time averages are equal to invariant 
distribution averages. The invariant distribution is identified as the 
distribution of the phase $P(\theta)$ in the $N\rightarrow\infty$ limit. In this 
limit the average of the scattering matrix $\langle S\rangle$, the optical $S$ 
matrix is equal to the fixed point value $S_\infty$. This was verified by using 
the fixed point solution in the expression of Poisson's kernel for $P(\theta)$. 
The theoretical distribution fits perfectly the histogram obtained from 
numerical iterations for a chain of delta potentials. The same is valid when the 
optical $S$ matrix is obtained from the numerical average of the scattering 
matrix. 

\section{Acknowledgements}

This work was supported by DGAPA-UNAM under project IN103115. We thank Centro 
Internacional de Ciencias for the facilities given for several group meetings 
and gatherings celebrated there. VD-R thanks the financial support of DGAPA. A. 
Robledo acknowledges support from DGAPA-UNAM-IN104417. We thank G. B\'aez, 
J.~A. M\'endez-Berm\'udez, and A.~M. Mart\'{\i}nez-Arg\"uello for useful 
comments.

\end{document}